# Rapid growth in a dual AGN during a gas-rich merger at z~4.5

**Hyewon Suh[1], Roberto Decarli[2], Emanuele Paolo Farina[1,2], Julia Scharwächter[1], Mar Mezcua[3,4], Giorgio Lanzuisi[2], Federica Loiacono[2], Brian C. Lemaux[1], Sukyoung K. Yi[5], Stefano Marchesi[6,7,2], Marta Volonteri[8], Günther Hasinger[9,10,11], Francesca Civano[12], Anniek Gloudemans[13], Ena Choi[14], Yeonwoo Nam[14], Adi Foord[15], Silvia Onorato[1]**

[1] International Gemini Observatory/NSF NOIRLab, 670 N. A'ohoku Place, Hilo, HI 96720, USA
[2] INAF - Osservatorio di Astrofisica e Scienza dello Spazio di Bologna, via Gobetti 93/3, 40129, Bologna, Italy
[3] Institute of Space Sciences (ICE, CSIC), Campus UAB, Carrer de Magrans, 08193 Barcelona, Spain
[4] Institut d'Estudis Espacials de Catalunya (IEEC), Edifici RDIT, Campus UPC, 08860 Castelldefels (Barcelona), Spain
[5] Department of Astronomy and Yonsei University Observatory, Yonsei University, Seoul 03722, Republic of Korea
[6] Dipartimento di Fisica e Astronomia "Augusto Righi", Università di Bologna, via Gobetti 93/2, I-40129 Bologna, Italy
[7] Department of Physics and Astronomy, Clemson University, Kinard Lab of Physics, Clemson, SC 29634, USA
[8] Institut d'Astrophysique de Paris (UMR 7095: CNRS & Sorbonne Universite), 98 bis Bd. Arago, F-75014, Paris, France
[9] TU Dresden, Institute of Nuclear and Particle Physics, 01062 Dresden, Germany
[10] DESY, Notkestraβe 85, 22607 Hamburg, Germany
[11] Deutsches Zentrum für Astrophysik, Postplatz 1, 02826 Görlitz, Germany
[12] NASA Goddard Space Flight Center, Greenbelt, MD 20771, USA
[13] NSF NOIRLab, Gemini Observatory, 670 N. A'ohoku Place, Hilo, HI 96720, USA
[14] Department of Physics, University of Seoul, 163 Seoulsiripdaero, Dongdaemun-gu, Seoul 02504, Republic of Korea
[15] Department of Physics, University of Maryland Baltimore County, 1000 Hilltop Cir, Baltimore, MD 21250, USA

**The late stages of galaxy mergers – when two supermassive black holes (SMBHs) reach kiloparsec-scale separations – represent a critical phase for understanding SMBH growth, galaxy co-evolution, and the progenitors of low-frequency gravitational waves. Yet confirmed close-separation (< 3 kpc) dual active galactic nuclei (AGNs) remain confined to the local Universe, despite predictions that such systems should be common in the merger-rich early Universe. Here we report the discovery of LID-1166, a dual AGN with a**

**projected separation of ~1.5 kpc at z~4.5, representing the first such system known beyond the local Universe. Spatially resolved JWST/NIRSpec integral-field spectroscopy reveals two spatially and kinematically distinct narrow-line emission components with a line-of-sight velocity offset of ~ -164 km/s. Both components additionally exhibit broad H$\alpha$ emission, including a compact off-nuclei broad H$\alpha$ component associated with the companion source, confirming two actively accreting SMBHs. Independent ALMA [CII] 158 μm observations reveal corresponding spatially and kinematically distinct cold gas components associated with the same nuclei, demonstrating that the system is a gas-rich merger. While undergoing super-Eddington accretion during a late-stage merger, the SMBHs already lie on the local black hole-host mass relation for massive elliptical galaxies within the uncertainties, suggesting that rapid, obscured growth may help establish this relation early in cosmic history. LID-1166 reveals a previously hidden phase of SMBH growth and points to a missing population of heavily obscured, merger-driven dual AGNs at high redshift.**

In the hierarchical structure formation paradigm, galaxies assemble through repeated mergers over cosmic time, a process expected to play a fundamental role in both stellar mass growth and the evolution of supermassive black holes[1,2] (SMBHs). Galaxy mergers can efficiently funnel large reservoirs of cold gas into galactic nuclei via tidal torques and shocks, triggering intense episodes of SMBH growth[3,4]. Theoretical models predict that such mergers naturally produce pairs of SMBHs and that gas-rich major mergers can induce short-lived phases of rapid, sometimes super-Eddington, accretion under heavy dust and gas obscuration[5,6]. These episodes offer a compelling pathway to alleviate the tension between SMBH seed formation models and the existence of over-massive black holes observed in the early Universe, as increasingly revealed by recent JWST observations[7,8].

Observationally, enhanced AGN activity has been reported in interacting and merging galaxies, particularly as systems approach coalescence[9-13]. Simulations predict that SMBH accretion peaks when black hole pairs reach separations below ~10 kpc, where merger-driven gas inflows are most effective[14]. The late-stage mergers, corresponding to SMBH separations of a few kpc or less, are therefore of particular interest, as they trace the final phase before the formation of gravitationally bound SMBH binaries[15,16]. However, this regime is expected to be heavily obscured by dense gas and dust, as demonstrated by studies of local late-stage mergers that host some of the most Compton-thick AGNs known[17-19]. These conditions severely hinder optical AGN identification and make close-separation dual AGNs intrinsically difficult to detect, especially at high redshift, where sub-arcsecond spatial resolution is required.

As a consequence, confirmed dual AGNs remain exceedingly rare[20,21]. Only a handful of systems with projected separations below ~7 kpc have been identified at $z>0.5$ (refs. 22-26), and all confirmed dual AGNs with separations below 3 kpc are confined to the local Universe[18]. Recent Gaia-based systematic searches have identified candidate dual quasars at z>1 (refs. 26-28), but these methods are biased toward luminous, unobscured optical quasars and have not revealed any < 3 kpc pairs. JWST has begun to uncover obscured AGNs and merging systems at z~>3 (refs. 29-33). Although close-separation dual AGN candidates have been proposed[7,34], no system has been unambiguously confirmed at separations < 3 kpc beyond the local Universe. This persistent absence stands in contrast to theoretical expectations, which predict that such systems– and the most rapid phases of SMBH growth– should be most common in the gas-rich, merger-dominated early Universe[35].

Here we report the discovery of a dual AGN, LID-1166, with a projected separation of ~1.5 kpc at z~4.5, representing the first close-separation dual AGN known beyond the local Universe and revealing heavily obscured, rapid black-hole growth in a late-stage galaxy merger. LID-1166 was identified as an X-ray-selected AGN in the *Chandra* COSMOS Legacy Survey[36] yet remains undetected even in the deepest *Hubble Space Telescope* (*HST*) optical and near-infrared imaging[37], suggesting that it is representative of a largely hidden black hole population. Its strong X-ray emission indicates powerful ongoing accretion, demonstrating that LID-1166 traces a distinct and previously unexplored phase of early SMBH growth, characterized by extreme obscuration coincident with intrinsically high accretion activity.

To identify the nature of LID-1166, we conducted spatially resolved integral-field spectroscopy with JWST NIRSpec/IFU. These observations reveal two compact nuclei in the H$\alpha$ emission-line region, separated by 0.23″, corresponding to a projected separation of ~1.5 kpc at z~4.5 (Fig. 1). After subtracting a model of the bright central AGN point spread function (PSF) generated using WebbPSF (see Methods), the PSF-subtracted H$\alpha$ channel maps clearly uncover a second compact nucleus offset from the primary source. The PSF-subtraction procedure and the resulting isolation of the companion AGN are illustrated in Extended Data Fig. 1. We extracted spatially resolved spectra for the two nuclei using circular apertures with radii of 0.2″. The primary nucleus spectrum was defined by the PSF model template derived from the original IFU cube, whereas the companion nucleus spectrum was extracted from the PSF-subtracted residual cube to minimize contamination from the primary AGN.

The resulting spectra reveal two spatially and kinematically distinct narrow-line H$\alpha$ components. Best-fitting systemic redshifts derived from the narrow-line H$\alpha$ emission indicate a line-of-sight velocity offset of ~ -164 +/- 3 km/s between the two nuclei, corresponding to redshifts z=4.502 and z=4.499. Both nuclei additionally exhibit broad H$\alpha$ emission characteristic of Type 1 AGNs, confirming that both sources host actively accreting SMBHs (Fig. 1). The broad-line emission displays distinct line profiles, widths, and centroid velocities. The primary component exhibits a narrow-line ratio of [NII]6583Å/H$\alpha$ = ~0.51 +/- 0.02, consistent with composite or AGN-like ionization, whereas the secondary component shows a lower ratio of ~0.13 +/- 0.08, more consistent with star-forming excitation or a lower-metallicity environment. The differing narrow-line ratios likely reflect differences in the local gas conditions and ionization environments of the two nuclei. The combination of spatially separated narrow-line regions with distinct kinematics, together with the detection of an off-nuclear broad-line component, demonstrates that LID-1166 hosts a genuine dual AGN in a late-stage galaxy merger.

The observed velocity offset and spectral differences robustly rule out gravitational lensing as an explanation for the double morphology. Multiple lensed images of a single background AGN must share the same intrinsic redshift and emission-line ratios, aside from modest microlensing-induced distortions of the broad-line profiles[38]. Moreover, the measured redshift difference ($\Delta z \sim 0.003$), the distinct broad H$\alpha$ line profiles, and differences in narrow-line diagnostics cannot be explained by AGN variability combined with lensing time delays, which are typically days to weeks[39]. These independent spectroscopic signatures provide decisive evidence that the two nuclei are physically distinct, actively accreting SMBHs rather than lensed images of a single source[40].

We estimate black hole masses for the primary and companion nuclei using the broad H$\alpha$ emission, finding log $M_{BH}/M_{\odot} \sim 7.8$ and 7.2, respectively (see Methods). Because the two nuclei are unresolved in X-rays, we derive the bolometric luminosity $L_{bol}$ for the combined system of both nuclei using the absorption-corrected X-ray luminosity and compute a system-integrated Eddington ratio by comparing $L_{bol}$ to the combined Eddington luminosity $L_{Edd}(M_{primary}+M_{companion})$. This yields an Eddington ratio of 3.7 [-1.3 +5.2], indicating super-Eddington accretion. The combination of spatially resolved broad-line AGN signatures, distinct kinematics, and extreme accretion rate demonstrates that LID-1166 hosts a rapidly growing dual SMBH system embedded in a heavily obscured merger at a small nuclear separation, at early cosmic times.

High-resolution ALMA observations of the [CII] 158 μm emission line provide an independent, kinematic confirmation of the dual AGN nature of LID-1166. The [CII] line, which arises from singly ionized carbon and traces multiple phases of the interstellar medium (ISM) largely independent of dust obscuration, is a robust probe of gas kinematics in high-redshift galaxies[41-43]. In merging systems, enhanced [CII] emission is often associated with gas-rich regions experiencing elevated star formation or disturbed, high-density gas driven by tidal interactions, making it a sensitive tracer of merger-driven ISM dynamics[42,44].

The [CII] emission in LID-1166 reveals two spatially and kinematically distinct velocity components associated with the primary and companion AGN positions identified from the JWST NIRSpec/IFU data, with velocity offsets consistent with those measured from the H$\alpha$ emission. We constructed continuum-subtracted, velocity-integrated [CII] intensity maps by integrating over a 1 GHz bandwidth (Fig. 2, bottom left). The integrated [CII] emission peak is spatially coincident with the primary AGN identified in the JWST data, while the overall line profile extracted from a circular aperture of radius 1″ centered on the primary AGN position shows a complex, non-Gaussian structure. Fitting the integrated spectrum with two Gaussian components yields velocity centroids at $z$=4.5028 and $z$=4.4989, consistent with the systemic redshifts independently measured from the JWST NIRSpec spectra.

Channel maps further reveal the spatial separation of the two kinematic components. At velocities near $\Delta v \sim -200$ km/s relative to the systemic redshift (defined at $z$=4.502), the [CII] emission peaks at the companion AGN position with separate intensity peaks detected at $> 6\sigma$ significance (Fig. 2, middle left). The velocity-offset component is spatially coincident with the companion AGN position identified in the JWST data. Extracting a spectrum at this location using a smaller aperture of radius 0.2″ yields a single Gaussian profile consistent with the lower-redshift component identified in the 1″-aperture spectrum, confirming that the two velocity components arise from spatially distinct regions. While [CII] emission can in some cases trace AGN-driven outflows[45,46], particularly through broad high-velocity wings, the [CII] components in LID-1166 are relatively narrow and spatially coherent. Their velocity offsets and spatial distribution are consistent with the two components independently identified in the JWST/NIRSpec data, supporting an interpretation in which the [CII] emission traces two interacting gas-rich components. We note, however, that some contribution from spatially extended outflow emission cannot be fully excluded.

Notably, the [CII] emission associated with the companion AGN is brighter than that of the primary nucleus. This enhanced [CII] emission likely reflects a larger cold-gas reservoir or more efficient gas excitation in the companion component, consistent with an asymmetric, gas-rich merger. Such behavior is inconsistent with expectations for a strongly lensed quasar, for which gravitational lensing is achromatic and would preserve flux ratios across wavelengths for images of the same background source. While modest chromatic effects can arise from differential magnification of emission regions with different spatial extents, such effects cannot account for the observed reversal in relative brightness between the two components. Together with the distinct velocities and spatial offsets, this provides further evidence against a lensing interpretation.

The total [CII] luminosity of the system is $L_{[CII]}=2.7\times10^{9}\ L_{\odot}$, yielding a molecular gas mass of $M_{H2}\sim8.2\times10^{10}\ M_{\odot}$ (see Methods). This gas reservoir is comparable to those observed in optically luminous quasars at high redshift[42,47] and is consistent with expectations for a gas-rich major merger. Despite being heavily obscured in the optical and hosting relatively modest SMBH masses, LID-1166 exhibits bright [CII] emission, demonstrating that the system contains a substantial cold gas supply capable of sustaining rapid black hole growth.

Although the dynamical mass of LID-1166 is uncertain owing to its disturbed kinematics and ongoing merger, we estimate a total dynamical mass of $M_{dyn}\sim5.1\times10^{10}\ M_{\odot}$ (see Methods). We emphasize that both the molecular gas mass and dynamical mass estimates are subject to systematic uncertainties, arising from the use of [CII] as a gas tracer and from projection effects in a non-virialized system. Nevertheless, even under conservative assumptions for the [CII]-to-$H_2$ conversion factor, LID-1166 remains a massive, gas-rich system in a late stage of merging, with a substantial fraction of its mass residing in cold molecular gas. This is consistent with theoretical expectations that dual AGNs preferentially reside in gas-rich halos compared to typical single AGN hosts[48].

Assuming that the dynamical mass provides a reasonable proxy for the bulge mass, LID-1166 is consistent with the locally derived black hole-host bulge mass relation for inactive, massive elliptical galaxies (Fig. 3; ref. 49). Because the dynamical mass includes contributions from stars, gas, and dark matter, it is expected to exceed the stellar bulge mass alone; accounting for this would shift LID-1166 modestly leftward in Fig. 3, further reinforcing its consistency with the local relation. Here we use *host mass* to refer generically to the mass of the galaxy hosting the SMBHs; in this system, the dynamical mass provides the only direct constraint and likely represents an upper bound on the bulge mass.

Our comparison is intentionally anchored to the ref. 49 relation, which is defined by inactive, massive elliptical galaxies widely interpreted as the end products of gas-rich major mergers. In contrast, local AGN relations[50] primarily probe actively accreting systems that may evolve through secular processes or minor mergers. Several high-redshift AGNs reported to host apparently overmassive black holes appear significantly offset relative to local $M_{BH}$-$M_{stellar}$ relations for AGNs, yet are much less discrepant when compared to the ref. 49 relation, as illustrated in Fig. 3. LID-1166, observed during an ongoing major merger at z~4.5, already lies on the ref. 49 relation when using its dynamical mass as a proxy for bulge mass. This suggests that, even during a phase of rapid, merger-driven growth, the global mass scale of the system is comparable to that of the massive elliptical galaxies that define the local scaling relation, implying that sufficient host mass is already assembled at this early epoch.

To investigate the expected evolution of LID-1166, we estimate the black hole mass accretion and host-galaxy star formation. Assuming a radiative efficiency of $\varepsilon$=0.1, the AGN bolometric luminosity yields a black hole mass accretion rate (BHAR) of ~ 6.3 $M_\odot$/yr. The host galaxy star formation rate (SFR) is estimated to be ~ 68 $M_\odot$/yr using the empirical [CII]-SFR calibration[51]. This is consistent with gas-rich, star-forming galaxies at z~4 but below the extreme starburst levels often observed in the most luminous quasars. We note that [CII]-based SFR estimates in AGN host galaxies remain uncertain because AGN radiation fields, shocks, and possible [CII] suppression in the central regions may affect the observed luminosity. Nevertheless, the narrow/core [CII] emission is generally understood to trace the host galaxy ISM and correlate with star formation activity over a broad range of redshifts. Using the [CII]-based molecular gas mass estimate ($M_{H2}$~$8.2\times10^{10}$ $M_\odot$), we infer a gas depletion timescale of ~1.2 Gyr, indicating that the observed level of star formation could plausibly the sustained over much longer timescales than the current AGN phase.

The evolutionary trajectories projected over 10 Myr and 100 Myr (Fig. 3) assume that the currently inferred BHAR and SFR remain constant, and therefore should be interpreted as illustrative upper limits on the possible growth during the observed phase rather than as predictive evolutionary tracks. In reality, SMBH accretion is likely highly episodic, and the duration of the current super-Eddington phase may be substantially shorter than the star formation timescale. Under this assumption, the trajectories indicate preferentially rapid SMBH growth relative to the host galaxy. Despite this accelerated growth, the system remains broadly consistent with the local scaling relation within the uncertainties. This suggests that the observed super-Eddington accretion phase is likely transient and episodic, consistent with theoretical and observational frameworks in which rapid SMBH assembly occurs through short-lived duty cycles.

Placed in this context, LID-1166 may represent a snapshot of rapid SMBH growth during a major merger that has not driven the system far from the scaling relation observed in local massive galaxies. While accreting at super-Eddington rates, the black holes in LID-1166 lie on the local black hole-host mass relation within the uncertainties, indicating that intense, obscured accretion episodes triggered by mergers can rapidly build black hole mass while preserving the global co-evolution between SMBHs and their hosts. In this picture, merger-driven super-Eddington accretion represents a brief but efficient growth phase, such that its short duty cycle allows SMBHs to grow rapidly without becoming over-massive relative to their hosts. If subsequent gas consumption and dynamical relaxation proceed as expected, systems like LID-1166 may plausibly evolve into massive elliptical galaxies in the local Universe, with both black hole and host masses largely in place by early cosmic times.

LID-1166 provides a compelling example of an actively accreting dual AGN in a close-separation merger during cosmic morning ($z\sim4.5$), filling a long-standing gap between theoretical expectations and observations of dual AGNs at high redshift. Hierarchical merger models predict that such systems should be common during early cosmic epochs, yet they have remained largely undetected. The inferred gas fraction and dynamical mass of LID-1166 are comparable to those of luminous quasars at $z\sim>6$, supporting the interpretation that it is an actively assembling system undergoing gas-rich mergers during a critical phase of galaxy and SMBH growth. Recent JWST studies have begun to reveal spatially extended H$\alpha$ emission around rapidly accreting black holes at high redshifts[37], hinting that mergers may be more prevalent than previously recognized even when stellar components remain undetected.

The presence of a dual AGN at a projected separation of only ~1.5 kpc is particularly remarkable. This separation corresponds to the merger stage at which simulations predict extreme, obscured, and potentially super-Eddington accretion to be most efficient[5,14]. Despite exhibiting broad H$\alpha$ emission characteristic of Type 1 AGNs, LID-1166 remains undetected in deep rest-frame UV/optical imaging and shows strong X-ray obscuration. This indicates that compact nuclear obscuration and larger-scale host-galaxy dust attenuation coexist, consistent with expectations for gas-rich merger-driven SMBH growth. Given its high bolometric luminosity (log $L_{bol}$~46.6 erg/s), the system is not intrinsically faint but instead heavily attenuated at rest-frame UV/optical wavelengths, consistent with a deeply embedded phase of rapid SMBH growth.

Although the current data do not provide a clean detection of two spatially separated stellar continuum components, the interpretation of LID-1166 as an interacting dual AGN system is supported by the combination of two spatially and kinematically distinct narrow-line regions, a compact off-nuclear broad-line component associated with the companion source, and corresponding spatially and kinematically distinct [CII] components resolved by ALMA. Notably, LID-1166 was not selected as a dual AGN candidate a priori, but emerged from a parent population of optically invisible X-ray sources identified in the *Chandra* COSMOS Legacy Survey. These systems exhibit strong X-ray emission despite extremely faint rest-frame UV/optical emission, indicating heavily obscured accretion activity. LID-1166 therefore highlights a fundamental observational limitation: close-separation dual AGNs in late-stage mergers may evade detection in broad-band imaging surveys because their host galaxies are heavily obscured and effectively invisible at rest-frame UV/optical wavelengths. The merger in LID-1166 is revealed only through spatially resolved AGN emission lines, suggesting that a substantial population of compact, obscured dual AGNs may remain hidden at high redshift.

The confirmation of LID-1166 as a close-separation dual AGN therefore provides a clear proof of concept for the power of high-resolution, spatially resolved spectroscopy in uncovering a hidden population of dual AGNs at high redshift. IFU observations are uniquely capable of isolating nuclear emission lines even when the underlying host galaxies are invisible. Our results suggest that close-separation (< 3 kpc) dual AGNs may have been missed in previous surveys owing to heavy obscuration, thereby biasing current census estimates. Systematic IFU studies with JWST, complemented by adaptive-optics (AO)-assisted near-infrared IFU spectroscopy from the ground (i.e., Gemini/GNIRS, Gemini/GIRMOS, Keck/OSIRIS, VLT/ERIS) and future 30-m class telescopes, will be essential for revealing this missing population and establishing the true role of mergers in driving rapid SMBH growth across cosmic time.

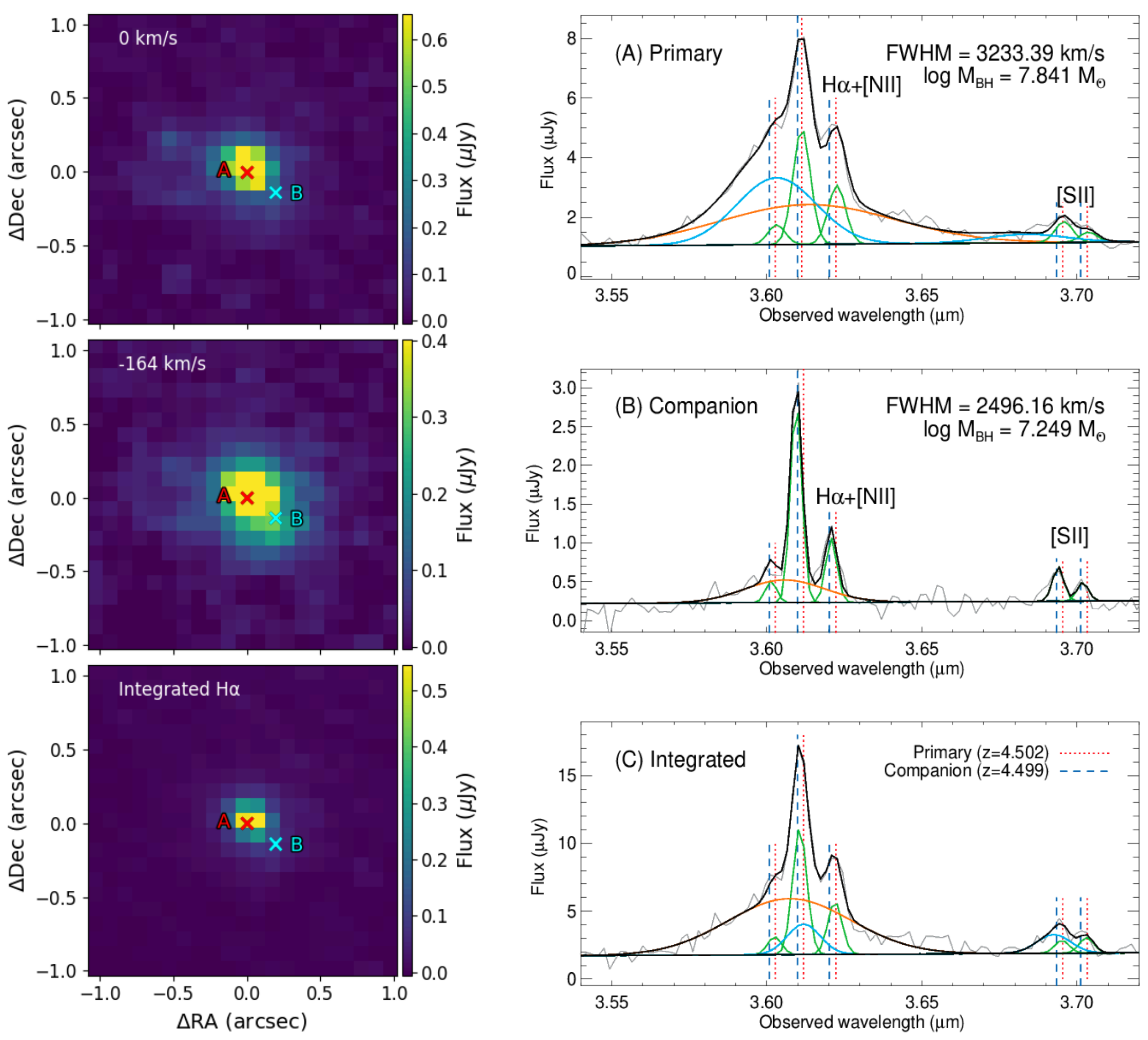

**Fig. 1 | JWST NIRSpec/IFU detection of a close-separation dual AGN in LID-1166.**
(Left) H$\alpha$ emission line maps from JWST NIRSpec/IFU observations. Top: Velocity channel map at 0 km/s (defined at z=4.502). Middle: Velocity channel map at -164 km/s, corresponding to the velocity offset of the companion nucleus. Bottom: Velocity-integrated map over the H$\alpha$+[NII] region. Red and cyan crosses mark the positions of the primary (A) and companion (B) nuclei, determined from centroiding of the PSF-subtracted H$\alpha$ emission (see Extended Data Fig. 1). The projected separation of the two nuclei is 0.23″ (~1.5 kpc at z~4.5).
(Right) Spatially resolved H$\alpha$ spectral fitting. The primary spectrum (A) corresponds to the PSF model template extracted from a circular aperture of radius 0.2″ centered on the primary nucleus in the original IFU cube. The companion spectrum (B) was extracted from the PSF-subtracted residual cube within a 0.2″ radius aperture centered on the companion nucleus. Spectrum (C) shows the integrated emission extracted within 0.6″ radius aperture. Grey curves show the observed spectra, and black curves show the best-fit models, which include a power-law continuum (black), narrow-line components (green), broad-line components (orange and blue). Vertical red dotted and blue dashed lines indicate the systemic redshifts of the primary (z=4.502) and companion (z=4.499), respectively. Both nuclei exhibit distinct broad H$\alpha$ emission, confirming two actively accreting SMBHs.

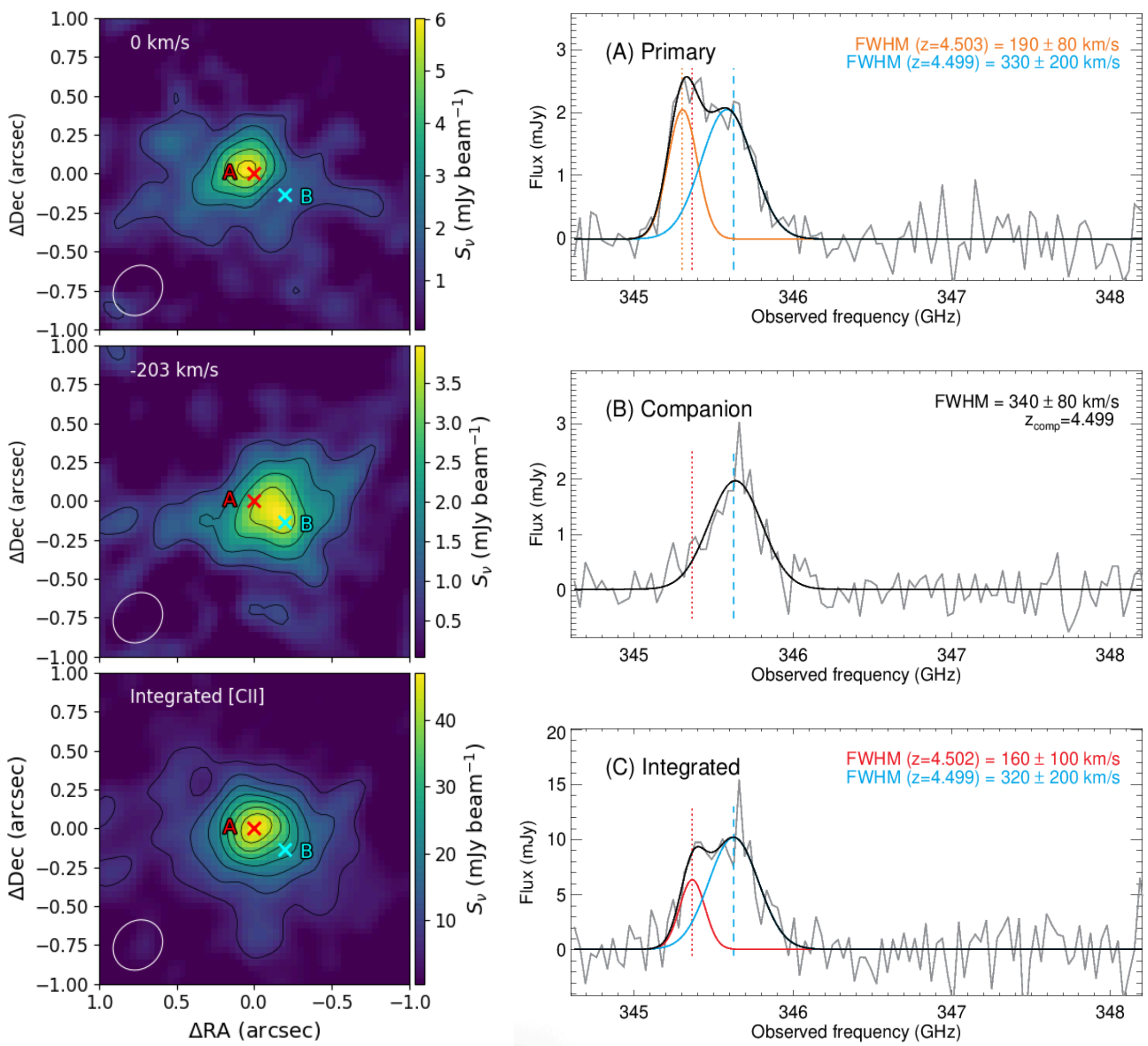

**Fig. 2 | ALMA [CII] emission reveals two kinematically distinct components in LID-1166.** (Left) Continuum-subtracted ALMA [CII] 158 μm maps. Top: Velocity channel map at 0 km/s (defined at z=4.502). Middle: Velocity channel map at -203 km/s, where the velocity-offset companion component is apparent. Bottom: Velocity-integrated [CII] intensity map integrated over 1 GHz. Red and cyan crosses mark the positions of the primary (A) and companion (B)

AGNs identified from the JWST data. Black contours start at 2$\sigma$ and increase in steps of 2$\sigma$ in [CII] flux density. The synthesized beam is shown in the lower-left corner of each panel.
(Right) [CII] spectra extracted within circular apertures of radius 0.2″ for (A) the primary nucleus and (B) the companion nucleus, and within a 1″ radius for (C) the integrated emission. Grey curves show the observed spectra and black curves the best-fitting models. The primary and integrated spectra are best described by two Gaussian components with centroids at z=4.502 (4.503) and z=4.499, consistent with the systemic redshifts derived from JWST/NIRSpec H$\alpha$ emission. Vertical red (orange) dotted and blue dashed lines mark the expected [CII] line centers at these redshifts. The velocity-offset component is spatially coincident with the companion nucleus and detected at 6$\sigma$ significance, confirming that the two [CII] components arise from distinct, kinematically separated regions.

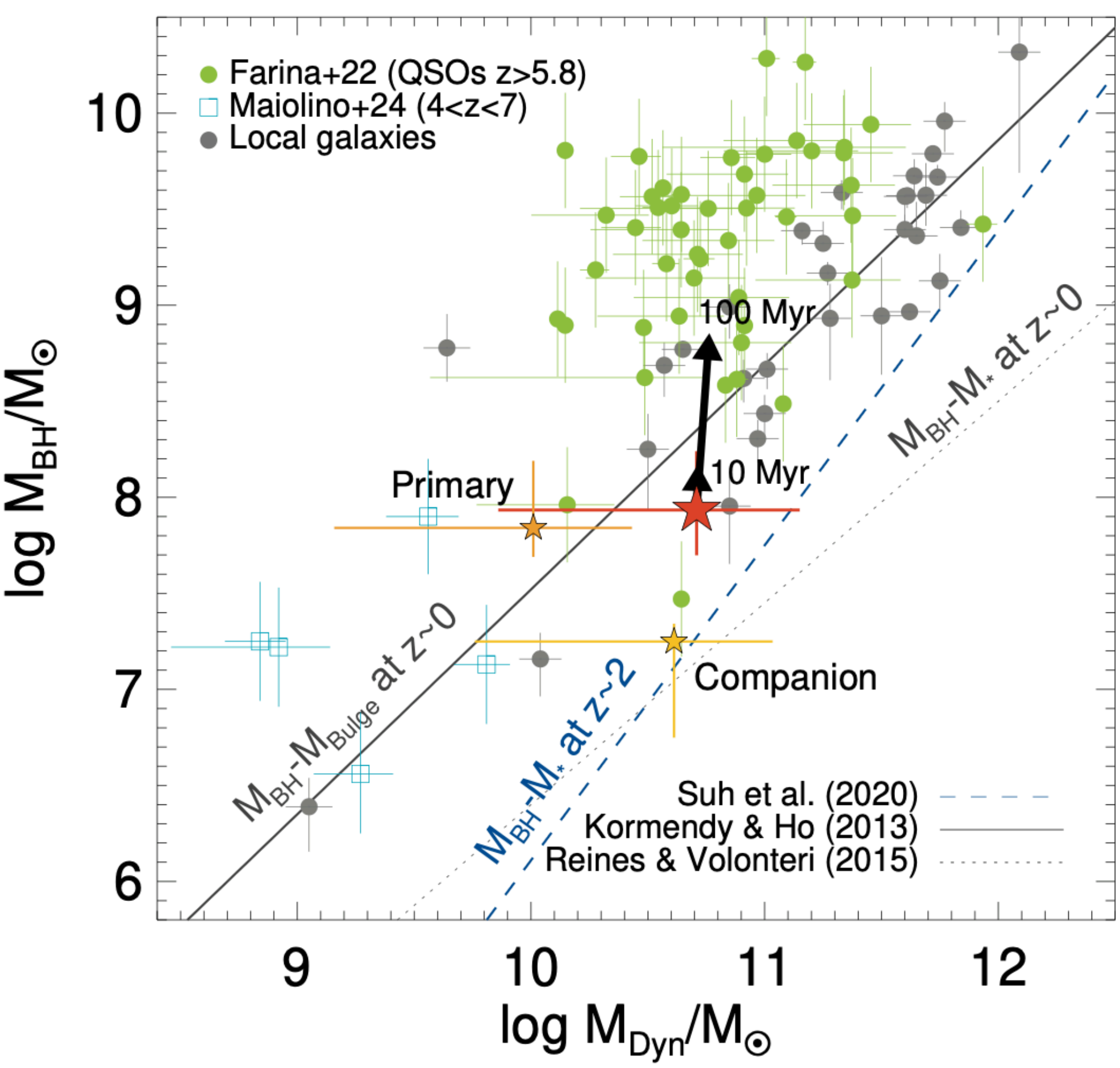

**Fig. 3 | Black hole mass versus dynamical mass.**
Black hole mass ($M_{BH}$) as a function of host galaxy dynamical mass ($M_{Dyn}$). LID-1166 is shown as a red star, representing the system-integrated values, while the primary and companion AGNs are shown as smaller orange and yellow stars, respectively (see Methods). UV-selected quasars at

z>5.8 (green circles; ref. 52) and less luminous AGNs at z~4-7 identified with JWST (blue squares; ref. 7) are shown for comparison. The local black hole–bulge mass for inactive, massive ellipticals (grey points) and its best-fitting relation (solid grey line) are taken from ref. 47. The blue dashed line shows the relation between black hole mass and total stellar mass for moderate-luminosity X-ray AGNs at z~1-2.5 (ref. 53), while the grey dotted line denotes the local black hole-stellar mass relation for AGNs from ref. 48. Within the uncertainties, LID-1166 is consistent with the local black hole-host mass scaling relation. The black arrows illustrate the expected evolution of LID-1166 assuming the inferred black hole accretion rate ($\dot{M}_{BH}$ ~6.3$M_{\odot}$/yr) and the star formation rate ($SFR_{[CII]}$~68 $M_{\odot}$/yr) remain constant for 10 Myr (shorter arrow) and 100 Myr (longer arrow). The evolutionary trajectory indicates more rapid SMBH growth relative to host galaxy growth, although the system remains broadly consistent with the local scaling relation within the uncertainties.

# Methods

## JWST NIRSpec/IFU PSF subtraction

We obtained JWST NIRSpec/IFU observations of LID-1166 as part of the Cycle 1 GO program 1760 (PI: Suh), using the G395M/F290LP grating/filter combination. This configuration provides spectral coverage from 2.9-5.1 µm at an average spectral resolution of R~1000 over a 3″x3″ field of view. The observations were carried out in April 2023 using the NRSIRS2 readout mode with a 4-point medium cycling dither, yielding a total on-source exposure time of 1.45 h.

Data reduction was performed with the JWST Science Calibration Pipeline (version 1.11.4) using the CRDS context jwst_1149.pmap. In addition to the standard pipeline processing, we applied custom steps to mitigate low-frequency 1/f noise, flag detector artifacts, reject residual outliers, and perform channel-by-channel background subtraction, following the procedure described in ref. 54.

To identify multiple nuclear components, we performed a point-spread-function (PSF) subtraction analysis on the fully calibrated NIRSpec IFU data. We extracted an integrated AGN spectrum from a circular aperture of radius 0.2″ (r=2 pixels) centered on the brightest spaxel and used this spectrum as a wavelength-resolved AGN PSF template. We assumed that the spectral shape of the PSF template is constant across the field of view, while its spatial amplitude variation follows the instrumental PSF. A model PSF was generated using WebbPSF and aligned to the data with sub-pixel accuracy via Fourier shifting. The simulated PSF was then scaled to match the BLR flux in each spaxel and subtracted from the data cube. This procedure removes the dominant central AGN emission while preserving residual emission from any spatially offset source. Any contamination of the PSF template by negligible host-galaxy light would lead to mild oversubtraction and therefore cannot artificially generate a spatially offset emission component.

In the PSF-subtracted Hα+[NII] cube, we detect a second compact emission component offset by ~0.23″ (corresponding to ~1.5 kpc at z~4.5) from the primary nucleus and exhibiting a relative line-of-sight velocity of ~ -164 +/- 3 km/s. This component is detected exclusively in the Hα emission line region and is absent in the adjacent continuum maps (see Extended Data Fig.1). The primary nucleus is centered at z=4.502, while the companion is detected at z=4.499. We note that spatially extended outflow emission from the primary nucleus may contribute to emission around the secondary nucleus. However, this extended residual emission is morphologically distinct from the compact emission peak associated with the secondary component. In particular, the narrow-line emission in the spatially integrated spectrum is shifted toward the velocity offset of the secondary component, indicating the secondary narrow-line region contributes significantly to the resolved ionized gas structure and may be spatially extended.
We verified the robustness of the detection through multiple tests, including shifting the PSF centroid, varying the extraction aperture used to construct the PSF template, and repeating the analysis with an independent calibration-star PSF. In all cases, the secondary compact source persists at the same spatial location and velocity offset, confirming that it is not an artifact of the PSF modelling and represents a genuine off-nuclear AGN. Extended Data Fig. 1 illustrates the PSF-subtraction procedure applied to the JWST NIRSpec/IFU data of LID-1166.

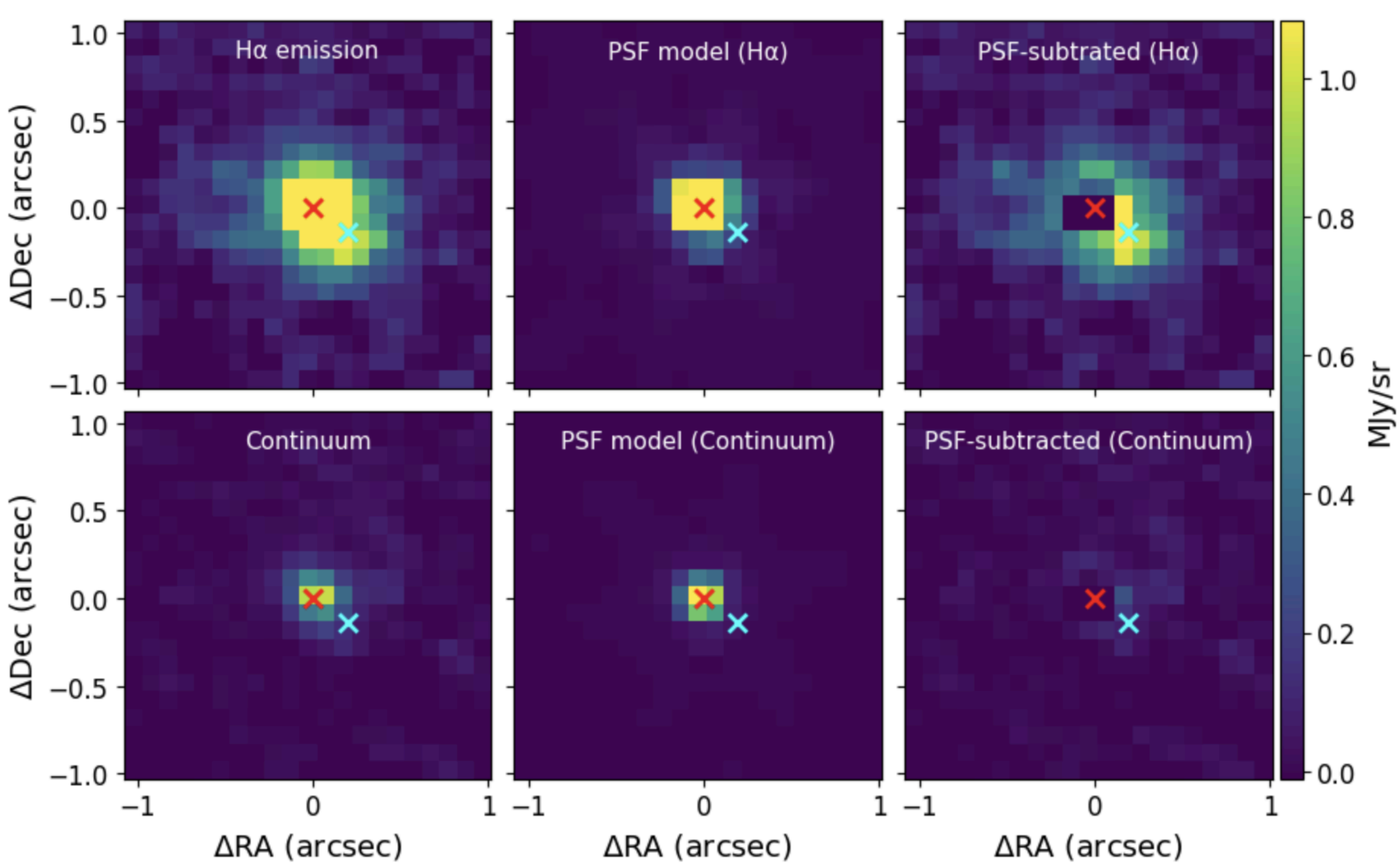

**Extended Data Fig. 1: PSF-subtraction of JWST NIRSpec/IFU maps revealing a dual AGN in LID-1166.**

Top panels show maps of the H$\alpha$+[NII] emission line region, while bottom panels show integrated continuum maps extracted from wavelength regions adjacent to the broad H$\alpha$ emission. From left to right, the panels display the original data cube, the best-fit PSF model generated using WebbPSF, and the residual image after PSF subtraction (original - PSF model), which isolates the companion AGN. Red and cyan crosses mark the positions of the primary and companion AGN, respectively.

## Black hole mass estimates

After PSF subtraction, we confirm the presence of two AGNs separated by 0.23″, corresponding to a projected separation of ~1.5 kpc at z~4.5, with both nuclei exhibiting broad H$\alpha$ emission. One-dimensional spectra were extracted using circular apertures of radius 0.2″ centered on each AGN position. The primary nucleus spectrum corresponds to the PSF model template derived from the original cube, whereas the companion spectrum was extracted from the PSF-subtracted residual cube to minimize contamination from the primary source.

Black hole masses were estimated using the single-epoch virial method based on the broad H$\alpha$ emission line width and luminosity. Emission-line fitting was performed on the continuum-subtracted spectra using a power-law model for the underlying continuum and multiple Gaussian components to represent both narrow and broad emission lines. Best-fit parameters were obtained via Levenberg-Marquardt least-squares minimization. The narrow-line region was modeled with Gaussian components for H$\alpha$ 6563 Å, [NII] 6548, 6583 Å, and [SII] 6716, 6731 Å. The flux ratio of the [NII] 6548, 6583 Å doublet was fixed at the theoretical value of 2.96, and the line widths and relative velocity offsets of all narrow components were tied to those of the narrow H$\alpha$ line.

The broad H$\alpha$ emission was modeled using one or two broad Gaussian components to best characterize asymmetries in the line profile, reflecting complex kinematics within the broad-line region. Broad Gaussian components were also allowed for the [SII] 6716, 6731 Å doublet when required by the data. The broad-line profile of LID-1166 shows significantly more complex and asymmetric structure than the relatively smooth broad-line profiles commonly observed in many low-redshift AGNs, which are often adequately described by a single broad component or modest asymmetry. In contrast, the line profiles in LID-1166 require multiple kinematic Gaussian components with noticeable velocity offsets from the systemic velocity. Such complex structures are more naturally interpreted as dynamically disturbed broad-line region gas associated with the merging or interacting nature of the system. This behavior also differs from the smooth extended broad H$\alpha$ wings observed in some "*little red dots*" at z>4, where Thomson scattering has been proposed as a possible explanation for the broad wings (ref. 55). The distinctly multi-component and asymmetric profiles observed in LID-1166 are inconsistent with a single smooth electron-scattering-dominated profile. We therefore conservatively treat all broad line components as part of the broad-line region for the black hole mass estimation.

The broad H$\alpha$ full width at half maximum (FWHM) and line luminosity were measured from the best-fit models and used to estimate black hole masses following the calibration of ref. 56. Measurement uncertainties were estimated using 1000 Monte Carlo realizations, in which the observed spectra were perturbed according to their flux uncertainties and refitted using the same procedure. For the primary nucleus, the two broad H$\alpha$ components have FWHM values of 5420 +/- 310 km/s and 2429 +/- 235 km/s, corresponding to an effective combined broad-line FWHM of 3233 +/- 222 km/s. The companion nucleus has a broad H$\alpha$ FWHM of 2494 +/- 105 km/s. Relative to the systemic velocity defined by the narrow H$\alpha$ emission, the two broad H$\alpha$ components of the primary nucleus have velocity offsets of -114 km/s and +22 km/s, while the broad H$\alpha$ component of the companion nucleus has a velocity offset of -213 km/s. Statistical uncertainties on the inferred black hole masses are small (~0.1 dex); however, systematic uncertainties associated with single-epoch virial mass calibrations introduce an intrinsic scatter of ~0.4 dex. Accounting for both statistical and systematic contributions, we estimate black hole masses of 6.9 [-2.0 +8.6]$\times 10^7$ $M_\odot$ for the primary AGN and 1.7 [-0.6 +0.5]$\times 10^7$ $M_\odot$ for the companion AGN.

## AGN Bolometric luminosity

Because the projected separation of the two nuclei (~0.2″) is well below the spatial resolution of the *Chandra* X-ray Observatory (~2″), the individual X-ray contributions of the two AGNs cannot be resolved[36]. We therefore estimate the AGN bolometric luminosity for the system as a whole using the absorption-corrected X-ray luminosity.

To account for the complex absorption and reflection processes expected in the case of heavy obscuration (e.g., Compton-thick; $N_H > 10^{24}$ cm$^{-2}$), we derived the intrinsic X-ray luminosity in the 2-10 keV using the MYtorus model[57,58]. This model self-consistently includes three components: (i) line-of-sight obscuration with Compton scattering applied to the intrinsic power-law continuum, (ii) a reflection component from the toroidal structure, and (iii) fluorescence emission line complex lines of Iron (Fe) and Nickel (Ni). The relative strengths of these components were fixed to be equal, corresponding to the default MYtorus configuration with a covering factor of 0.5 and a uniform column density. The inclination angle between the line-of-sight and the axis of the torus was set to 75°, ensuring interception of the obscuring material. We adopted a power-law photon index of $\Gamma$=1.9, consistent with typical AGN continua.

The choice of a Compton-thick (CT) obscuration model is motivated by the X-ray spectral properties themselves rather than by the optical/UV non-detection. Relative to a simple unabsorbed power-law fit, the MYtorus model provides an improved fit to the *Chandra* spectrum ($\Delta$C=3.38 for one additional free parameter). In addition, allowing the photon index to vary freely in a simple power-law fit yields an unusually flat continuum ($\Gamma$=1.07 [+0.70 -0.67]), significantly harder than the canonical AGN value of $\Gamma$~1.9. Such flat observed spectra are commonly interpreted as signatures of heavy obscuration, where the direct continuum is suppressed and the observed emission becomes dominated by reflected and scattered components[58].

From the MYtorus modelling, we derive a hydrogen column density of $N_H$=1.0 [-0.8 +3.0] x $10^{24}$ $cm^{-2}$, formally Compton-thick although the lower bound extends into the heavily obscured Compton-thin regime, and an absorption-corrected rest-frame 2–10 keV luminosity of $L_{2\text{-}10\ keV}$=6.7 [-4.4 +5.1]x$10^{44}$ erg $s^{-1}$. The AGN bolometric luminosity is then estimated by applying a luminosity-dependent bolometric correction following ref. 59, yielding log $L_{bol}$=46.6 [-0.8 +0.5] erg $s^{-1}$. We verified that the qualitative interpretation of LID-1166 remains unchanged under simple non-CT spectral models. In those cases, the inferred bolometric luminosity (log $L_{bol}$~45.3-45.7) and Eddington ratio (0.2-0.5) decrease, but the system still remains a luminous, rapidly accreting dual AGN embedded in a gas-rich late-stage merger.

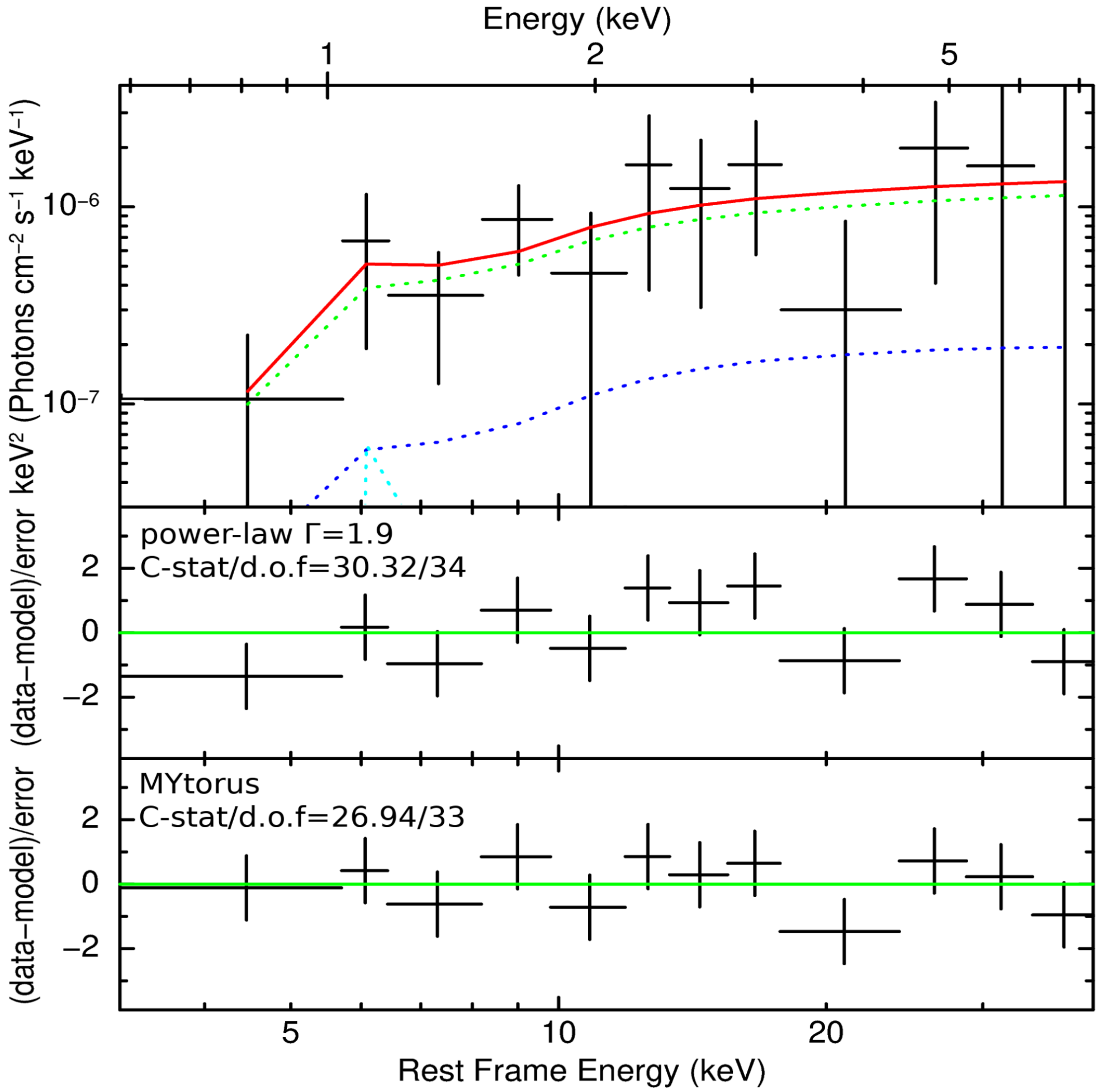

**Extended Data Fig. 2: X-ray spectral modeling of the LID-1166 system.**

Rest-frame unfolded *Chandra* spectrum of LID-1166 fitted with the MYtorus model. In the upper panel, the total best-fitting model is shown in red, together with the primary absorbed continuum (green), reflected/scattered emission component (blue), and fluorescent emission component (cyan). The middle and bottom panels show the data-to-model residuals for the simple power-law fit and the MYtorus fit, respectively. The corresponding observed-frame energy range is indicated on the upper x-axis.

**ALMA [CII] observations**

We obtained Atacama Large Millimeter/submillimeter Array (ALMA) observations of the [CII] 158 μm line in LID-1166 under program ID 2024.1.01025.S (PI: Suh). The observations were carried out on March 16, 2025, with a total on-source integration time of 24.65 minutes and a synthesized beam of ~0.2″, providing sub-kpc spatial resolution at z~4.5. The observations were tuned to a central frequency of 345.544 GHz, corresponding to the redshifted [CII] transition (rest frequency 1900.5369 GHz) at the systemic redshift of the source at z=4.5.

The data were calibrated using the standard ALMA calibration pipeline with CASA (version 6.6.1.17, ref. 60), following the procedure described in Decarli et al. (2018). The continuum was subtracted from the visibility plane using the CASA task UVCONTSUB, enabling the construction of continuum-subtracted [CII] spectral cubes. The final data cube was generated with a velocity channel width of 30 km/s and reaches an rms sensitivity of 0.05 mJy/beam per channel. All subsequent measurements and analyses were carried out on primary-beam-corrected images.

We constructed the [CII] velocity and velocity dispersion maps from the continuum-subtracted [CII] cube (Extended Data Fig. 2). The velocity field exhibits a coherent gradient across the two nuclei, spanning approximately -300 to +200 km/s relative to the systemic redshift defined by the primary nucleus. The kinematic axis appears broadly aligned with the projected separation of the two nuclei, linking the gas dynamics to the ongoing merger geometry. The [CII] velocity dispersion peaks near the integrated flux centroid, likely reflecting unresolved velocity structure and beam smearing within the synthesized beam. To further examine the kinematic structure, we extracted a position-velocity (PV) diagram along the line connecting the two nuclei (Extended Data Fig. 2, right). The PV diagram reveals a continuous velocity gradient across the system, with spatially offset emission components associated with each nucleus. The emission is not consistent with a single symmetric rotating disk centered on one nucleus; instead it shows kinematically distinct components connected by intermediate-velocity gas. This structure supports a scenario in which the [CII]-emitting gas traces two interacting systems embedded within a disturbed and dynamically evolved gas distribution, consistent with an ongoing late-stage merger. Overall, the [CII] kinematics indicate two spatially and kinematically distinct components embedded in a disturbed gas reservoir, reinforcing the interpretation of LID-1166 as a merging dual AGN system at z~4.5.

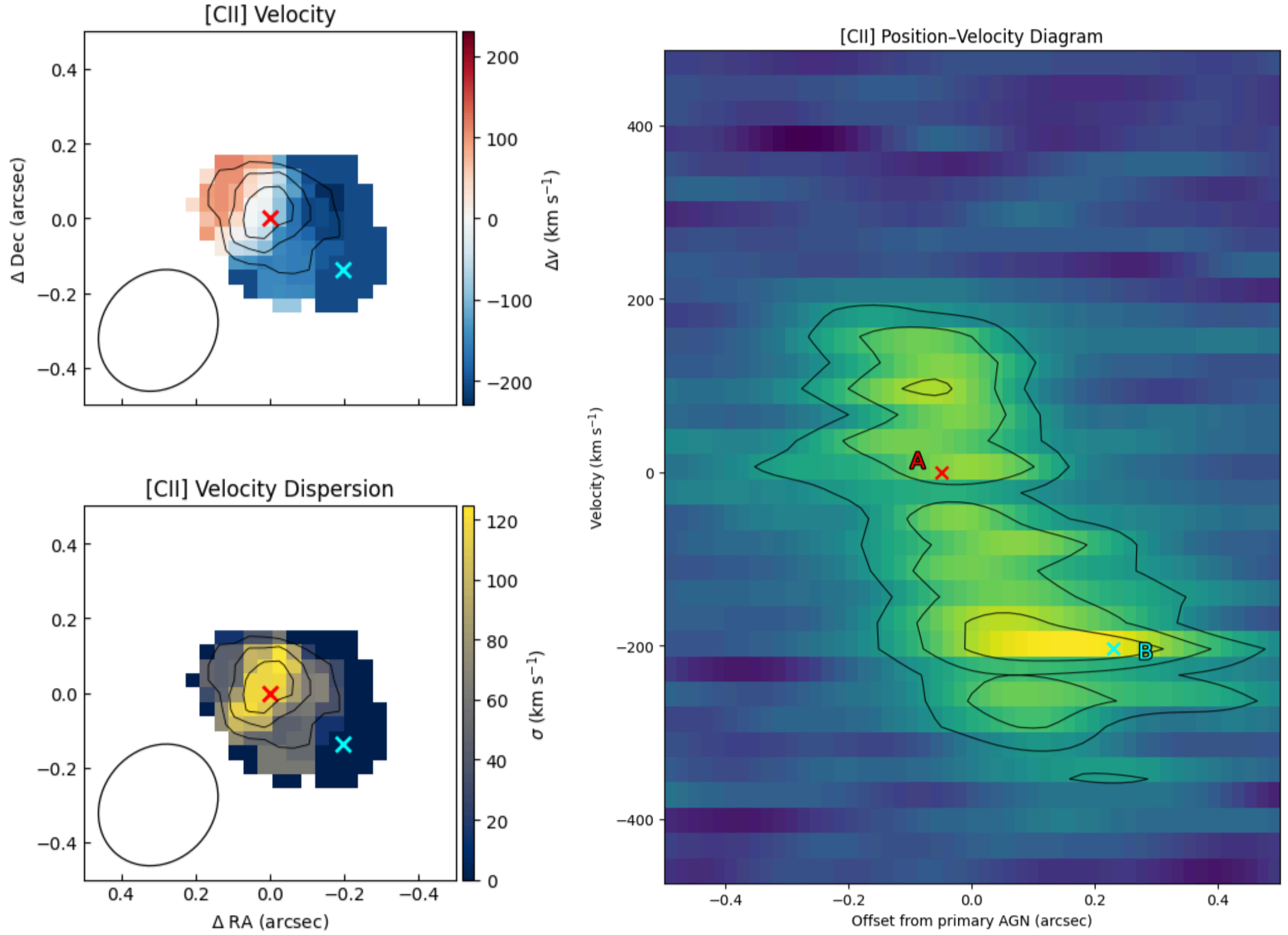

**Extended Data Fig. 3: ALMA [CII] velocity and velocity dispersion maps of LID-1166.**
Top left: [CII] velocity map of the continuum-subtracted [CII] 158 µm emission, shown relative to the systemic redshift z=4.502 defined by the primary nucleus. A coherent velocity gradient is observed across the system, spanning several hundred km/s. Bottom left: [CII] velocity dispersion map. The velocity dispersion peaks near the integrated flux centroid and decreases toward larger radii. Right: Position-velocity (PV) diagram extracted along the line connecting the two nuclei, with spatial offset measured relative to the primary nucleus. Red and cyan crosses mark the positions of the primary and companion AGNs identified from the JWST/NIRSpec data. In the left panels, black contours traces the [CII] flux density, starting at $2\sigma$ and increasing in steps of $2\sigma$. In the PV diagram, contours are shown at $2\sigma$, $3\sigma$, and $4\sigma$. The synthesized beam is shown in the lower-left corner of the left panels.

## Molecular gas mass from [CII]

The [CII] 158 µm fine-structure line is a widely used tracer of the cold interstellar medium (ISM) in high-redshift galaxies and provides an empirical means to estimate the molecular gas mass[41,42.61,62]. We use the ALMA [CII] observations of LID-1166 to quantify the molecular gas content of the system.

To measure the total [CII] emission, we extracted a spectrum centered on the primary AGN position identified from the JWST/NIRSpec IFU data using a circular aperture with a radius of 1″. The aperture size was determined using aperture-growth measurements, in which we measured the enclosed [CII] flux within progressively larger apertures and adopted the radius at which the enclosed flux approximately converged. The final 1″ aperture encompasses both AGN components and captures the bulk of the [CII] emission from the system. The extracted spectrum shows significant departures from a single Gaussian profile, indicating dynamically complex gas kinematics consistent with an ongoing merger. We therefore modeled the spectrum with two Gaussian components. The total [CII] flux within the 1″ aperture is 4.5 +/- 2.0 Jy km/s. The redshifts of the two velocity components are consistent with those independently measured from the H$\alpha$ emission in the JWST/NIRSpec IFU data.

We also extracted [CII] spectra at the positions of the primary and secondary AGNs using smaller apertures with radii of 0.2″. The spectrum extracted at the position of the primary AGN is best described by two Gaussian components, with a total [CII] flux of 1.1 +/- 0.6 Jy km/s. The spectrum extracted at the position of the secondary AGN is well described by a single Gaussian component with an integrated flux of 0.7 +/- 0.2 Jy km/s and a FWHM of 340 +/- 80 km/s, centered at z=4.498, consistent with the blueshifted component identified in the 1″-aperture spectrum. Together, these measurements indicate that the [CII] emission traces spatially and kinematically distinct gas components associated with the two AGNs.

We compute the [CII] luminosities following Equation (1) of ref. 41, obtaining $L_{[CII]}=6.7\times10^8\ L_\odot$ for the primary component, $2.1\times10^9\ L_\odot$ for the secondary component, and a total $L_{[CII]}=2.7\times10^9\ L_\odot$ for the system. We estimate the molecular gas mass using a [CII]-to-$H_2$ conversion factor of $\alpha_{[CII]}\sim 30\ M_\odot/L_\odot$ (ref. 61), yielding a total molecular gas mass of $M_{H2}\sim8.2\times10^{10}\ M_\odot$. We note that [CII]-based molecular gas mass estimates may be systematically higher by a factor of ~3-5 compared to those derived from other traces[44,63], and that $\alpha_{[CII]}$ can span a wide range (~10-1000) depending on star-formation surface density and gas depletion time[64]. Even accounting for these uncertainties, LID-1166 hosts a massive cold gas reservoir of order $\sim10^{10}M_\odot$, consistent with expectations for a gas-rich major merger at high redshift.

Although [CII] emission can, in principle, receive contributions from AGN-heated gas, studies of high-redshift systems indicate that the dominant [CII] luminosity is typically associated with the host galaxy ISM, and primarily trace the distribution and kinematics of cold gas[44,65]. Any AGN-related contribution to the [CII] emission would bias the inferred gas mass high, rendering our estimates conservative upper limits.

## Dynamical mass estimates

Estimating the dynamical mass of LID-1166 is complicated by its disturbed kinematics and merging nature, which violate the assumptions of a relaxed, rotationally supported disk. We therefore adopt an empirical approach appropriate for high-redshift systems with complex gas dynamics. The resulting dynamical mass estimates should be interpreted as order-of-magnitude constraints rather than precise measurements.

We estimate the dynamical mass using the prescription given in Equation (14) of ref. 44, which relates the [CII] line width and spatial extent to the total enclosed mass and is calibrated using a sample of high-redshift quasars with resolved [CII] emission. Applying this method to the [CII] kinematic measurements, we derive a total dynamical mass of $M_{dyn}$~$5.1\times10^{10}$ $M_{\odot}$. When applied separately to the two kinematic components, this yields dynamical masses of $1.0\times10^{10}$ $M_{\odot}$ and $4.1\times10^{10}$ $M_{\odot}$ for the primary and secondary components, respectively.
The apparent discrepancy between the dynamical and molecular gas masses likely arises from differences in spatial scales and underlying assumptions. The dynamical mass is derived from the central [CII] kinematics and assumes a simplified geometry, whereas the molecular gas mass may trace a more extended component. In a merger system, disturbed and non-virialized motions can further bias the dynamical estimate low.

**Data availability**

The data for ALMA and JWST used in this study are publicly available through their respective data archives. These observations are associated with the JWST GO program #1760 and the ALMA program #2014.1.01025.S. Other data generated and/or analyzed during the study are available from the corresponding author upon reasonable request.

**Acknowledgements**
HS, EPF, JS, BCL, AG, SO are supported by the international Gemini Observatory, a program of NSF NOIRLab, which is managed by the Association of Universities for Research in Astronomy (AURA) under a cooperative agreement with the National Science Foundation, on behalf of the Gemini partnership of Argentina, Brazil, Canada, Chile, the Republic of Korea, and the United States of America. S.K.Y. acknowledges support from the Korean National Research Foundation (RS-2025-00514475 and RS-2022-NR070872). EC is supported by the National Research Foundation of Korea (NRF-RS-2025-00515276).
This work is based on observations made with the NASA/ESA/CSA James Webb Space Telescope. The data were obtained from the Mikulski Archive for Space Telescopes at the Space Telescope Science Institute, which is operated by the Association of Universities for Research in Astronomy, Inc., under NASA contract NAS 5-03127 for JWST. These observations are associated with program #1760.
Support for program #1760 was provided by NASA through a grant from the Space Telescope Science Institute, which is operated by the Association of Universities for Research in Astronomy, Inc., under NASA contract NAS 5-03127.
This paper makes use of the following ALMA data: ADS/JAO.ALMA#2024.1.01025.S. ALMA is a partnership of ESO (representing its member states), NSF (USA) and NINS (Japan), together with NRC (Canada), MOST and ASIAA (Taiwan), and KASI (Republic of Korea), in cooperation with the Republic of Chile. The Joint ALMA Observatory is operated by ESO, AUI/NRAO and NAOJ. The National Radio Astronomy Observatory is a facility of the National Science Foundation operated under cooperative agreement by Associated Universities, Inc.

**Author Contributions**
H.S. is the Principal Investigator of the JWST and ALMA proposals, led the analysis and interpretation of the results, and drafted the manuscript. R.D reduced and analyzed the ALMA data. F.L reduced the JWST NIRSpec/IFU data. S.M. provided *Chandra* X-ray spectra. G.L. analyzed the X-ray data and wrote the relevant section. S.K.Y., E.C., and Y.N. performed simulations and provided discussions on black hole growth. E.P.F., J.S., M.M., B.C.L., M.V., G.H., F.C., A.G., A.F., S.O. helped with the interpretation of the results and provided comments on the analysis. All authors contributed to the discussion of the presented results and the preparation of the manuscript.